\documentclass{article}
\usepackage[english]{babel}
\usepackage{amsmath}
\usepackage{bm}
\usepackage{booktabs}
\usepackage{multirow}
\usepackage{url}
\usepackage{xcolor}
\usepackage{graphicx}
\usepackage{float}

\newtheorem{theorem}{Theorem}

\title{Smoothing spline density estimation from doubly truncated data}

\author{David Bamio  \\
		Instituto Español de Oceanografía (IEO - CSIC),\\
		Centro Oceanográfico de Vigo \\
		\and 
		Jacobo de Uña-Álvarez$^*$ \\
		Department of Statistics and Operations Research,\\
		Universidade de Vigo, \& CITMAga \\
		\and
		{\small $^*$Corresponding author. E-mail: jacobo@uvigo.gal}
	}
\begin{document}
	
	\maketitle

\begin{abstract}
	In Astronomy, Survival Analysis and Epidemiology, among many other fields, doubly truncated data often appear. Double truncation generally induces a sampling bias, so ordinary estimators may be inconsistent.	In this paper, smoothing spline density estimation from doubly truncated data is investigated. For this purpose, an appropriate correction of the penalized likelihood that accounts for the sampling bias is considered. The theoretical properties of the estimator are discussed, and its practical performance is evaluated through simulations. Two real datasets are analyzed using the proposed method for illustrative purposes. Comparison to kernel density smoothing is included. 
\end{abstract}
	
	\section{Introduction}
	
	Let us consider independent and identically distributed (iid) observations $\{X_i\}_{i=1}^n$ from some probability density $f$ on a bounded domain $\mathcal{X}$. We are interested in the nonparametric estimation of $f$ from the available sample. A density function $f$ has the intrinsic constraints of positivity ($f\ge 0$) and unity ($\int_{\mathcal{X}}f(x)dx=1$), so the obtained estimates must satisfy these constraints. \cite{good} proposed to carry the estimation of $f$ by minimizing the penalized likelihood functional
	\begin{equation}\label{gen-pen-lik}
		L(f)+\lambda J(f).
	\end{equation}
	Here, $L(f)$ is a goodness of fit measure, usually taken as the minus log-likelihood of the data; $J(\eta)$ is a roughness penalty defined by some notion of smoothness for functions in the domain $\mathcal{X}$, and $\lambda$ is the smoothing parameter that controls the trade-off between the two opposing goals. Minimization of (\ref{gen-pen-lik}) takes place in $\mathcal{H}=\{f\colon J(f)<\infty\}$ and the minimizer is called a \emph{smoothing spline}.
	
	Assuming $f>0$ on $\mathcal{X}$, one can use the logistic density transform $f(x)=e^{\eta(x)}/\int_{\mathcal{X}}e^{\eta}$ to estimate $\eta$ instead, which is free of the positivity and unity constraints \cite{leonard}. The penalized likelihood becomes
		\begin{equation}\label{l-den-est}
				-\frac{1}{n}\sum_{i=1}^{n}\eta(X_i)+\log\int_{\mathcal{X}}e^{\eta(x)}dx+\lambda J(\eta)
		\end{equation}
	and one may search for its minimizer in a reproducing kernel Hilbert space $\mathcal{H}$, with $J(\eta)$ being a square seminorm. However, one must note that $\eta$ is determined by $f$ only up to a constant; in order to make the logistic transform one-to-one and the first term in (\ref{l-den-est}) strictly convex, \cite{guqiu} proposed to enforce a side condition on the members of $\mathcal{H}=\{\eta\colon J(\eta)<\infty\}$ of the form $A\eta=0$, where $A$ is an averaging operator on $\mathcal{X}$; for instance, one may impose that $\int_{\mathcal{X}}\eta=0$, or $\eta(x_0)=0$ for a certain $x_0\in\mathcal{X}$. 
	Knowing that $L(\eta)$ is continuous and strictly convex in $\mathcal{H}=\{\eta\colon A\eta=0, J(\eta)<\infty\}$, if $L(\eta)$ has a minimizer (i.e., the maximum likelihood estimate) in the null space $\mathcal{N}_J=\{\eta\colon A\eta=0,J(\eta)=0\}$ then the existence and uniqueness of the minimizer of (\ref{l-den-est}) in $\mathcal{H}$ is guaranteed \cite[Th. 3.1]{guqiu}.
	
	In some settings an iid sample of the probability density of interest $f$ may not be available. In the case of biased sampling, the values of the target density are selected from the population with probabilities depending on the values themselves, so some corrections of the ordinary estimators are to be made in order to avoid a systematic estimation bias. Consider an iid sample $\{X_i\}_{i=1}^n$ on $\mathcal{X}$ from densities proportional to $w_i(x)f(x)$, where $w_i(x)\ge 0$ are known biasing functions. As in \cite[Section 7.6.1]{gu}, let $\mathcal{T}$ be an index set and $w(t,x)\in\mathcal{T}\times\mathcal{X}$ a function such that the set $\{w(t,\cdot), t\in\mathcal{T}\}$ includes all possible biasing functions with $w(t,\cdot)\neq w(t',\cdot)$ for $t\neq t'$. Within this setting, every biasing function $w_i$ can be written as $w(T_i,\cdot)$ for some $T_i\in\mathcal{T}$, and taking $T_i$ as observations from a probability density $m(t)\in\mathcal{T}$ one can treat the data as the pairs $(T_i,X_i)$ observed from a two-stage sampling process. Assuming $0<\int_{\mathcal{X}}w(t,x)f(x)dx<\infty$, the sampling conditional densities $w(t,x)f(x)/\int_{\mathcal{X}}w(t,\cdot)f$ of $X$ given $t$ are well defined. Using once again the logistic transform $f(x)=e^{\eta(x)}/\int_{\mathcal{X}}e^{\eta}$, the sampling conditional density of $X$ given $t$ becomes
	\begin{equation*}
		\frac{w(t,x)f(x)}{\int_{\mathcal{X}}w(t,\cdot)f}=\frac{w(t,x)e^{\eta(x)}}{\int_{\mathcal{X}}w(t,\cdot)e^{\eta}},\hspace{0.3 cm}x \in \mathcal{X},
	\end{equation*}
	leading to the penalized likelihood functional
	\begin{equation}\label{bias-pen-lik}
		-\frac{1}{n}\sum_{i=1}^{n}\left\{\eta(X_i)-\log\int_{\mathcal{X}}w(T_i,x)e^{\eta(x)}dx\right\}+\lambda J(\eta).
	\end{equation}
	
	\noindent When biased sampling arises from random truncation, the pairs $(T,X)$ are drawn from a joint density $g(t)f(x)$, $(t,x)\in\mathcal{T}\times\mathcal{X}$, but only the data inside the observable region $A\subset\mathcal{T}\times\mathcal{X}$ are recorded, while those in $A^c$ are lost. In this context, the biasing functions are $w(t,x)=I_{[(t,x)\in A]}$ and $m(t)\propto g(t)\int_{\mathcal{X}}I_{[(t,x)\in A]}f(x)dx$. For instance, for the standard model of independent left-truncation \cite{KM}, $g(t)$ is the density of the truncating variable and one has $A=\{(t,x)\in \mathcal{T} \times \mathcal{X}: t\leq x \}$.
	
	
	In Astronomy, Survival Analysis and Epidemiology, among many other fields, random double truncation may appear. The target variable $X$ is doubly truncated when there exist two random limits $(U,V)$ such that $X$ is observed only when $U\leq X\leq V$. \cite{quasars} found double truncation in the analysis of quasar luminosities. Epidemiological studies on event times, on their turn, often proceed through interval sampling. This means that the epidemiological registry is comprised of individuals who underwent the event of interest between two specific calendar times. Interval sampling results in random double truncation, which generally induces biased sampling. See \cite{dt.libro} for further details and examples.
	
Smoothing spline density estimation can be adapted to double truncation by extending the biasing function $w(t,x)$ to allow for a two-dimensional truncating argument $t=(u,v)$. Therefore, in principle, both the practical implementation and the theoretical validation of the estimator can be accomplished from existing software and results; see Sections \ref{sec:methods} and \ref{sec:sims} for the specific adaptations which are needed for such goals. Importantly, for the best of our knowledge the practical performance of smoothing splines when estimating a doubly truncated density has never been studied in the literature. The aim of this piece of work is to fill this gap and to compare the smoothing spline density estimator to the kernel density estimator, for which an adaptation to double truncation already exists; see \cite{jac-kde}. Other possible methods for estimating a density function under random double truncation involve parametric information on $X$ or $(U,V)$, so they are not considered in the present paper. This is the case of the semiparametric density estimator in \cite{jac-kde}, or the parametric setup in which $f$ is characterized by a finite number of parameters; see \cite{dorreemura} and references therein.

	The rest of the paper is organized as follows. In Section \ref{sec:methods} the methods to estimate a density function under double truncation are given, and their theoretical properties are discussed. A simulation study to explore the finite sample performance of the several methods is conducted in Section \ref{sec:sims}, where a comparison of smoothing splines and kernel density estimation is performed. Real data illustrations are given in Section \ref{sec:realdata}, while the main conclusions of this piece of work are reported in Section \ref{sec:discussion}.
	
	\section{Methods and theoretical properties}
	\label{sec:methods}
	
		\subsection{Smoothing spline density estimator}
	
	Let $\{(U_i,V_i,X_i)\}_{i=1}^n$ iid data with the conditional distribution of $(U,V,X)$ given $U\leq X\leq V$. It is assumed that $(U,V)$ is independent of $X$. The penalized likelihood functional (\ref{bias-pen-lik}) holds with $T_i=(U_i,V_i)$ and $w(u,v,x)=I_{[u\leq x\leq v]}$. Hence, a smoothing spline to consistently estimate $f$ under double truncation can be introduced as the minimizer of (\ref{bias-pen-lik}). Note that, in the notation of Section 1, one has $A=\{(u,v,x) \in \mathcal{T} \times \mathcal{X}:u\leq x\leq v\}$, so the index set $\mathcal{T}$ is a subset of the Euclidean two-dimensional space.
	
	An interesting feature of double truncation is that the sampling bias may be negligible. This occurs whenever the selection probability for a specific $X$-value $x$, $G(x)=P(U\leq X\leq V|X=x)=P(U\leq x\leq V)$, is constant along $x$. With interval sampling, for instance, it holds $U=V-\tau$ for some positive constant $\tau$ (the length of the sampling interval) and, hence, $G(x)=P(x\leq V \leq x+\tau)$. Thus, in the particular scenario of interval sampling, the density of the sampled $X_i$ coincides with the target $f$ when $V$ is uniformly distributed on a convenient support interval; see \cite{dua_sim} for further motivation. In such a case, one may proceed to estimate $f$ by minimizing the standard penalized likelihood (\ref{l-den-est}), which may entail a variance reduction compared to the minimizer of (\ref{bias-pen-lik}). This is explored through simulations in Section \ref{sec:sims}.

		\subsection{Kernel density estimator}

Kernel smoothing is a well-known approach to nonparametrically estimate a smooth density function $f$ from random samples \cite{WJ}. The standard kernel density estimator is defined as the convolution of a re-scaled kernel function $K_h(\cdot)=K(\cdot/h)/h$ and the empirical distribution function of the data $\{X_i\}_{i=1}^n$, given by $F^*_n(x)=n^{-1}\sum_{i=1}^n I_{[X_i\leq x]}$:

\begin{equation}\label{kde}
	f_h(x)=\int K_h(x-t)dF^*_n(t)=\frac{1}{n} \sum_{i=1}^n K_h(x-X_i).
\end{equation}

\noindent In (\ref{kde}), $K$ is a zero-mean density function and $h=h_n$ is a sequence of positive real numbers (called bandwidths, or smoothing parameters) with $h_n \rightarrow 0$ as $n \rightarrow \infty$.

A kernel density estimator is conceptually simpler than a smoothing spline. In particular, the kernel density estimator has a closed form. The trade-off between bias and variance in (\ref{kde}) is automatically controlled through the selection of an appropriate bandwidth parameter $h$. The estimator (\ref{kde}) can be easily extended to sampling scenarios different from random sampling; for this, one just replaces $F^*_n$ in (\ref{kde}) with a proper substitute.

\cite{jac-kde} introduced kernel density estimation for doubly truncated data as

\begin{equation}\label{kde_dt}
	f_h(x)=\int K_h(x-t)dF_n(t)
\end{equation}

\noindent where $F_n$ denotes the nonparametric maximum-likelihood estimator (NPMLE) of $F$, the cumulative distribution attached to $f$, under random double truncation. This estimator $F_n$ can be obtained as the maximizer of the conditional likelihood of the $X_i$ given the $(U_i,V_i)$,

\begin{equation}\label{lik_dt}
	\mathcal{L}_n(F)=\prod_{i=1}^n \frac{dF(X_i)}{F(V_i)-F(U_i-)},
\end{equation}

\noindent on the class of all the cumulative distributions. See \cite{quasars} for seminal discussions. Although (\ref{kde_dt}) is a natural estimator, it requires $F_n$, for which there is no explicit form and must be computed numerically. Also, \cite{xiao} showed that the existence and uniqueness of $F_n$ is not guaranteed. A possible solution is to impose some structure on $F$ before maximizing (\ref{lik_dt}). This is what smoothing splines do; note that, indeed, the minus log of (\ref{lik_dt}) is just the minus log-likelihood in (\ref{bias-pen-lik}). Another advantage of smoothing splines is that they do not suffer from the boundary effects that frequently occur in kernel smoothing. All these aspects are discussed using the simulation results in Section \ref{sec:sims}.

	\subsection{Existence of the estimators}
	
	Smoothing splines and kernel smoothing have different conditions for the existence and uniqueness of the density estimator. We will now describe these conditions starting with the smoothing spline density estimator, where we assume without loss of generality that the smoothing parameter for the penalized likelihood functional is $\lambda=1$.
	
	\begin{theorem}[\cite{guqiu}]\label{exist.spline}
		Suppose $L(\eta)$ is a continuous and strictly convex functional in a Hilbert space $\mathcal{H}$ and $J(\eta)$ is a square seminorm in $\mathcal{H}$ with a finite dimensional null space $\mathcal{N}_J\subset\mathcal{H}$. If $L(\eta)$ has a minimizer in $\mathcal{N}_J$, then $L(\eta)+J(\eta)$ has a unique minimizer in $\mathcal{H}$.
	\end{theorem}

	Given a doubly truncated dataset $(U_i,V_i,X_i)$, $i=1,\dots,n$, kernel density estimation is based on the NPMLE of the cumulative distribution function $F$. A necessary and sufficient condition for the existence and uniqueness of the NPMLE was provided by \cite{vardi}, and this condition can be checked graphically as shown by \cite{xiao}. Let us define a graph $\mathcal{G}$ with $n$ vertices, each of them representing an observation triplet $(U_i,V_i,X_i)$, such that there is a directed edge from vertex $i$ to vertex $j$ if and only if $X_j\in[U_i,V_i]$. A directed path is a sequence of edges connecting a sequence of distinct vertices with all the edges directed in the same direction. If, for any two vertices $i,j\in \mathcal{G}$, there exists a directed path from $i$ to $j$ and vice versa, then we say that the graph $\mathcal{G}$ is strongly connected. We are now in the position to give the two following theorems.

	\begin{theorem}[\cite{xiao}]\label{unique.npmle}
		There exists a unique NPMLE of $F$ if and only if $\mathcal{G}$ is strongly connected.
	\end{theorem}

	\begin{theorem}[\cite{xiao}]\label{exist.npmle}
		If $\mathcal{G}$ is connected but not strongly connected, then a NPMLE of $F$ does not exist.
	\end{theorem}
	
	\cite{xiao} provide an example of a doubly truncated dataset for which the NPMLE does not exist; the data are shown in Table~\ref{toyexample}. Since there is no $X_i$ that is contained in $[U_7,V_7]$ apart from $X_7$, there is no direct path from vertex 7 to any other vertex and thus the graph $\mathcal{G}$ is not strongly connected. The graph is nonetheless connected, and thus by Theorem~\ref{exist.npmle} the NPMLE does not exist. This implies in particular that the kernel density estimator (\ref{kde_dt}) is not well defined. However, minimization of (\ref{bias-pen-lik}) is feasible for this dataset. The intuitive explanation is that smoothing spline estimation introduces a structure in the target density (the Hilbert space attached to the penalty term) before optimization is performed.
	
	

	\begin{table}[H]
		\centering
		\begin{tabular}{@{}ccc@{}}
			\toprule
			$i$ & $X_i$ & $[U_i,V_i]$    \\ \midrule
			1   & 0.75  & $[0.4,2]$    \\
			2   & 1.05  & ${[}0.3,1.4{]}$  \\
			3   & 1.25  & ${[}0.8,1.8{]}$  \\
			4   & 1.5   & ${[}0,2.3{]}$    \\
			5   & 2.25  & ${[}1.3,2.6{]}$  \\
			6   & 2.4   & ${[}1.1,3{]}$   \\
			7   & 2.5   & ${[}2.45,3.4{]}$ \\ \bottomrule
		\end{tabular}
		\caption{Example of a doubly truncated dataset retrieved from \cite{xiao}.}
		\label{toyexample}
	\end{table}

	\subsection{Convergence of the estimators}
	
	For kernel density estimation, conditions ensuring convergence to the target were investigated in \cite{jac-kde}, see also \cite{quasars}. Such conditions are similar to those corresponding to random sampling, once the identifiability of $f$ has been ensured; this means that the sampling probability $G(x)$ must be positive along the support of $f$. The asymptotic bias of the kernel smoother, of order $h^2$, is just the one corresponding to random sampling. However, the asymptotic variance, with order $(nh)^{-1}$, depends on $G$, being relatively larger for $x$-points with a small sampling probability $G(x)$. See \cite{jac-kde} for details.
	
	The asymptotic properties of smoothing splines from biased data are collected in \cite{gu}. In particular, under certain conditions the in-probability order of the Kullback-Leibler distance between the estimator and the target is $n^{-1}\lambda^{-1/r}+\lambda^p$, where $r$ and $p$ depend on the smoothness of $f$ and $J(f)$; see \cite[Section 9.2]{gu}. Then, the role of $\lambda$ in smoothing splines is similar to that of $h$ for kernel smoothing: larger values of $\lambda$ will introduce more bias in order to reduce the variance. For biased sampling, high-level assumptions on the biasing function $w(t,x)$ for the asymptotic results to hold are given in \cite[Section 9.2.5]{gu}. Primitive conditions on the variables $X$ and $(U,V)$ that may ensure such high-level assumptions in the particular case of random double truncation are however missing in the literature.
	
	\section{Simulation study}
	\label{sec:sims}
	
	A series of simulations were conducted in order to test the effectiveness of the smoothing spline approach to density estimation from doubly truncated data. Along with the bias-corrected smoothing spline density estimator that minimizes (\ref{bias-pen-lik}), the ordinary smoothing spline method based on (\ref{l-den-est}) which ignores the potential sampling bias as well as the nonparametric kernel density estimator (KDE) described in \cite{jac-kde} were considered for comparison purposes. Four different settings for the distribution of observations $X$ and left-truncation limits $U$ were considered, which are
	
	\begin{itemize}
		\item S1. $X\sim \text{Unif}(0,1)$, $U\sim \text{Unif}(-1/3,1)$ (no sampling bias).
		\item S2. $X\sim \text{Unif}(0,1)$, $[(3/4)(U+1/3)]^{1/2}\sim \text{Unif}(0,1)$.
		\item S3. $X\sim \text{Beta}(3/2,5)$, $[(3/4)(U+1/3)]^{1/2}\sim \text{Unif}(0,1)$.
		\item S4. $X\sim \text{Norm}(1/2,1/10)$, $U\sim \text{Beta}(20,20)$.
	\end{itemize}
	Moreover, two different cases were considered for the right-truncation boundary $V$, obtained as $V=U+\tau$. The first case corresponds to an interval sampling context, where the observation interval length is constant; we take $\tau=1/3$. In the second case, some noise is introduced to the interval length as $\tau\sim \text{Unif}(1/3-1/20,1/3+1/20)$. Therefore, eight different settings were used for the simulation of $M=250$ trials $\{(U_i,V_i,X_i)\}_{i=1}^n$ of sizes $n=100$ and $n=200$. For each sample, density estimates from the three methods described above were obtained, producing a total of $48$ different scenarios.
	
	The ordinary smoothing spline estimates were computed using the function \texttt{ssden} function from the \texttt{gss} R package~\cite{gugss}. To compute the corrected smoothing spline estimates the function \texttt{ssden} was modified. Specifically, the \texttt{bias} argument of the function \texttt{ssden} was expanded to allow for three arguments. This is important since, for doubly truncated data, the sampling condition is given by $w(t,x,\tau)=I_{\{x\in[t,\,t+\tau]\}}$ where $t=U$ is the left-truncating value and $\tau=V-U$ is the length of the truncation interval, which may be random. Smoothing parameter selection was performed using the cross-validation method based on the Kullback-Leibler distance described by \cite[Section 7.3]{gu}, with $\alpha=1.4$ as suggested by the author. For kernel density estimates, the \texttt{densityDT} function from the \texttt{DTDA} R package~\cite{dtda} was used, employing the one-step direct plug-in rule for bandwidth selection, which is one of the selectors recommended by \cite{morvk}. This bandwidth was passed to the function \texttt{densityDT} by choosing the option \texttt{bw="DPI1"}. All estimates from the three methods were discretized on a grid consisting of 101 equally spaced points in the $[0,1]$ support.
	
\begin{figure}[H]
		\centering
		\includegraphics[width=\textwidth]{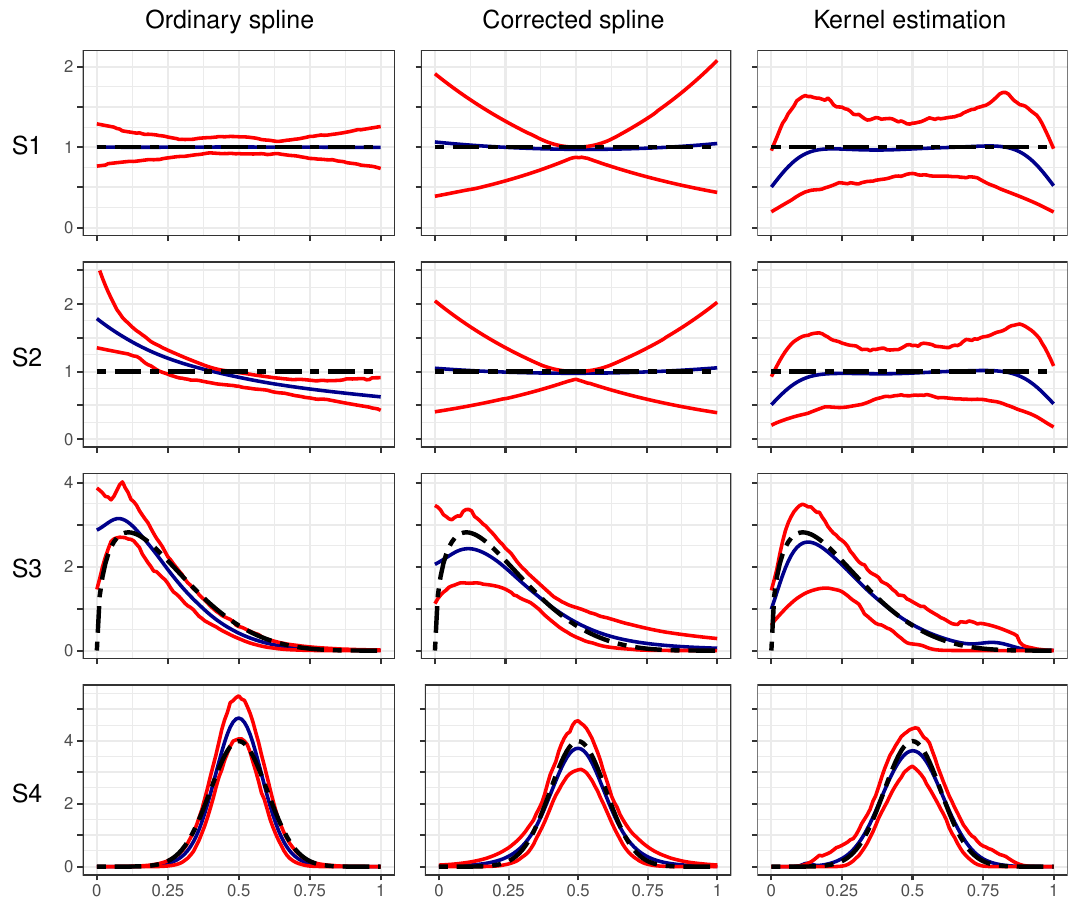}
		\caption{Density estimation results after 250 Montecarlo simulations of sample size $n=200$ for the the interval sampling settings (constant $\tau$). Represented in blue is the mean of the 250 density estimates obtained for each point in the support grid; in red, the 2.5\% top and bottom percentiles. The theoretical density is represented in a black dashed line.}
		\label{panel_is}
	\end{figure}

\begin{figure}[H]
		\centering
		\includegraphics[width=\textwidth]{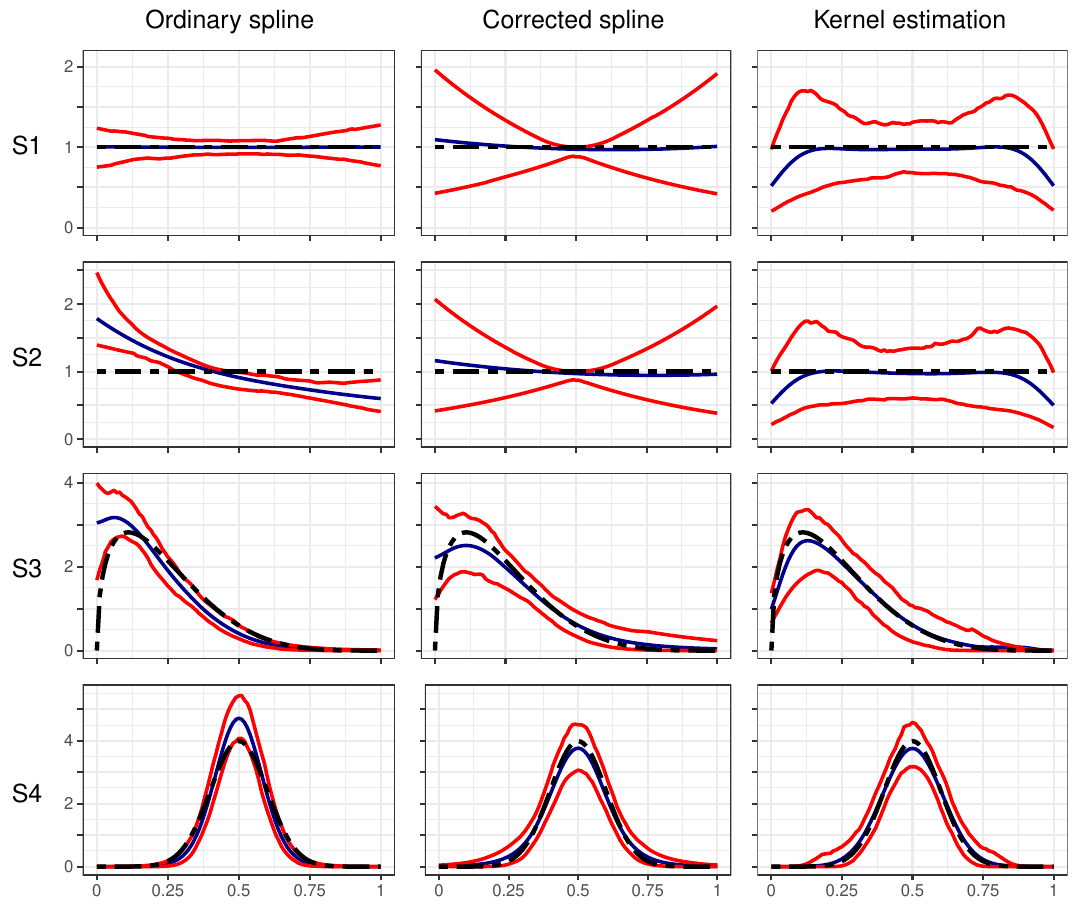}
		\caption{Density estimation results after 250 Montecarlo simulations of sample size $n=200$ for settings with random observation interval length $\tau$. Represented in blue is the mean of the 250 density estimates obtained for each point in the support grid; in red, the 2.5\% top and bottom percentiles. The theoretical density for each setting is represented in a black dashed line.}
		\label{panel_tau}
	\end{figure}

	Figures~\ref{panel_is} and~\ref{panel_tau} report the density estimates obtained for each method and simulation scenario with $n=200$, allowing for a visual comparison. The black dashed lines indicate the true densities, the blue solid lines represent the mean of the 250 estimates at each of the 101 points of the support grid, and the red solid lines correspond to the 2.5\% top and bottom percentiles across the 250 trials, representing 95\% oscillation limits. Regarding the S1 and S2 rows, it can be addressed that the corrected smoothing spline introduces some noise in the estimation, which increases the variability of the estimates compared to the ordinary smoothing spline. Interestingly, for both S1 and S2 the corrected spline confidence limits present a bow-tie shape, showing much less variability towards the centre of the support. In these two scenarios the KDE suffers from boundary effects, which may actually be advantageous in the S3 scenario and is otherwise negligible for S4.
	
	In order to evaluate the discrepancy between density estimate $\hat{f}$ and the true density $f$, the integrated squared error ISE($f,\hat{f}$)$=\int_{0}^{1}[f(x)-\hat{f}(x)]^2dx$ was obtained for each estimate. Table~\ref{tabla_mise} provides the mean (MISE) and standard deviation (SDISE) of the ISE for each simulated scenario. Except for the S1 case, where there is no sampling bias, the corrected smoothing spline performs similarly to the ordinary spline for $n=100$ samples, albeit with higher standard deviations, and shows significant improvements for $n=200$. Moreover, in every case except the S4 scenario, the KDE exhibits a MISE larger than that of the corrected spline estimator. This is influenced by the fact that some of the weights estimated by KDE are degenerate, leading to a very large value of ISE; this happens in S3 for both sample sizes and S4 for $n=100$, as can be easily guessed from the large standard deviations attached to the KDE. Such degeneration is linked to the aforementioned issues of the NPMLE for doubly truncated data, which may not exist or may not be unique; is such cases, the solution provided by the numerical algorithms may be useless.
	
	To perform a fair comparison that excludes degenerate estimations, Table~\ref{tabla_mediana} reports robust measures of the ISE, namely median (MDISE) and interquartile range (IQRMISE). With this approach, the comparison between the corrected spline and the KDE becomes more ambivalent: the former still performs better for S1 and S2, while KDE shows significant improvements for S3 and similar or marginally better performance for S4, generally exhibiting lower IQRMISE across all scenarios. These results appear to be largely explained by the influence of the boundary effects already observed in Figures \ref{panel_is} and \ref{panel_tau}.
	
	\begin{table}[H]
		\renewcommand{\arraystretch}{1.1}
		\centering
			\begin{tabular}{cc|ccc|ccc|}
				\cline{3-8}
				\multicolumn{1}{l}{}                                                                               & \multicolumn{1}{l|}{} & \multicolumn{3}{c|}{$n=100$}  & \multicolumn{3}{c|}{$n=200$}  \\ \hline
				\multicolumn{2}{|c|}{Model}                                                                                                & ORD     & COR    & KDE     & ORD     & COR    & KDE     \\ \hline
				\multicolumn{1}{|c|}{\multirow{8}{*}{\begin{tabular}[c]{@{}c@{}}Constant $\tau$\end{tabular}}} & \multirow{2}{*}{S1}   & .0117   & .1075   & .1446   & .0064   & .0578   & .0798   \\
				\multicolumn{1}{|c|}{}                                                                             &                       & (.0163) & (.1619) & (.1237) & (.0090) & (.0748) & (.0609) \\ \cline{2-8} 
				\multicolumn{1}{|c|}{}                                                                             & \multirow{2}{*}{S2}   & .1147   & .1138   & .1372   & .1105   & .0539   & .0844   \\
				\multicolumn{1}{|c|}{}                                                                             &                       & (.0787) & (.1635) & (.1247) & (.0521) & (.0731) & (.0670) \\ \cline{2-8} 
				\multicolumn{1}{|c|}{}                                                                             & \multirow{2}{*}{S3}   & .2531   & .2358   & .3332   & .1997   & .1377   & .2580   \\
				\multicolumn{1}{|c|}{}                                                                             &                       & (.1398) & (.1623) & (1.678) & (.0827) & (.0985) & (1.204) \\ \cline{2-8} 
				\multicolumn{1}{|c|}{}                                                                             & \multirow{2}{*}{S4}   & .0984   & .0993   & .0794   & .0905   & .0513   & .1338   \\
				\multicolumn{1}{|c|}{}                                                                             &                       & (.0813) & (.0808) & (.0617) & (.0556) & (.0396) & (.0651) \\ \hline
				\multicolumn{1}{|c|}{\multirow{8}{*}{Random $\tau$}}                                                      & \multirow{2}{*}{S1}   & .0142   & .0951   & .1348   & .0054   & .0535   & .0786   \\
				\multicolumn{1}{|c|}{}                                                                             &                       & (.0183) & (.1411) & (.1147) & (.0079) & (.0701) & (.0536) \\ \cline{2-8} 
				\multicolumn{1}{|c|}{}                                                                             & \multirow{2}{*}{S2}   & .1199   & .1255   & .1467   & .1189   & .0617   & .0914   \\
				\multicolumn{1}{|c|}{}                                                                             &                       & (.0760) & (.1643) & (.1312) & (.0551) & (.0786) & (.0734) \\ \cline{2-8} 
				\multicolumn{1}{|c|}{}                                                                             & \multirow{2}{*}{S3}   & .2772   & .2354   & .2616   & .2216   & .1281   & .1813   \\
				\multicolumn{1}{|c|}{}                                                                             &                       & (.1444) & (.1895) & (1.493) & (.0934) & (.0774) & (.9762) \\ \cline{2-8} 
				\multicolumn{1}{|c|}{}                                                                             & \multirow{2}{*}{S4}   & .1043   & .1037   & .0770   & .0891   & .0544   & .0466   \\
				\multicolumn{1}{|c|}{}                                                                             &                       & (.0841) & (.1156) & (1.295) & (.0561) & (.0442) & (.0288) \\ \hline
			\end{tabular}
		\caption{Mean ISE, and standard deviation of the ISE in brackets, of the several density estimators across 250 trials: ordinary smoothing spline (ORD), corrected smoothing spline (COR) and kernel density estimator (KDE). The length of the sampling interval is $\tau$ and $n$ is the sample size.}
		\label{tabla_mise}
	\end{table}

	\begin{table}[H]
		\renewcommand{\arraystretch}{1.1}
		\centering
		\begin{tabular}{cc|ccc|ccc|}
			\cline{3-8}
			\multicolumn{1}{l}{}                                                                               & \multicolumn{1}{l|}{} & \multicolumn{3}{c|}{$n=100$}  & \multicolumn{3}{c|}{$n=200$}  \\ \hline
			\multicolumn{2}{|c|}{Model}                                                                                                & ORD     & COR    & KDE     & ORD     & COR    & KDE     \\ \hline
			\multicolumn{1}{|c|}{\multirow{8}{*}{\begin{tabular}[c]{@{}c@{}}Constant $\tau$\end{tabular}}} & \multirow{2}{*}{S1}   & .0046   & .0458   & .1067   & .0029   & .0298   & .0604   \\
			\multicolumn{1}{|c|}{}                                                                             &                       & (.0157) & (.1330) & (.0575) & (.0079) & (.0758) & (.0575) \\ \cline{2-8} 
			\multicolumn{1}{|c|}{}                                                                             & \multirow{2}{*}{S2}   & .0936   & .0449   & .0950   & .0996   & .0240   & .0641   \\
			\multicolumn{1}{|c|}{}                                                                             &                       & (.0917) & (.1458) & (.1225) & (.0588) & (.0654) & (.0568) \\ \cline{2-8} 
			\multicolumn{1}{|c|}{}                                                                             & \multirow{2}{*}{S3}   & .2198   & .1889   & .0876   & .1878   & .1077   & .0622   \\
			\multicolumn{1}{|c|}{}                                                                             &                       & (.1797) & (.1929) & (.0997) & (.1130) & (.0856) & (.0649) \\ \cline{2-8} 
			\multicolumn{1}{|c|}{}                                                                             & \multirow{2}{*}{S4}   & .0807   & .0775   & .0643   & .0835   & .0402   & .0411   \\
			\multicolumn{1}{|c|}{}                                                                             &                       & (.0800) & (.1020) & (.0594) & (.0680) & (.0444) & (.0394) \\ \hline
			\multicolumn{1}{|c|}{\multirow{8}{*}{Random $\tau$}}                                                      & \multirow{2}{*}{S1}   & .0082   & .0489   & .0972   & .0025   & .0293   & .0618   \\
			\multicolumn{1}{|c|}{}                                                                             &                       & (.0156) & (.1012) & (.1231) & (.0064) & (.0603) & (.0588) \\ \cline{2-8} 
			\multicolumn{1}{|c|}{}                                                                             & \multirow{2}{*}{S2}   & .1079   & .0574   & .1034   & .1079   & .0296   & .0665   \\
			\multicolumn{1}{|c|}{}                                                                             &                       & (.1045) & (.1650) & (.1301) & (.0783) & (.0884) & (.0727) \\ \cline{2-8} 
			\multicolumn{1}{|c|}{}                                                                             & \multirow{2}{*}{S3}   & .2549   & .1794   & .0848   & .2033   & .1055   & .0585   \\
			\multicolumn{1}{|c|}{}                                                                             &                       & (.1933) & (.1869) & (.0875) & (.1261) & (.0845) & (.0482) \\ \cline{2-8} 
			\multicolumn{1}{|c|}{}                                                                             & \multirow{2}{*}{S4}   & .0787   & .0661   & .0607   & .0769   & .0436   & .0402   \\
			\multicolumn{1}{|c|}{}                                                                             &                       & (.0898) & (.0963) & (.0651) & (.0751) & (.0529) & (.0367) \\ \hline
		\end{tabular}
		\caption{Median ISE, and interquartile range of the ISE in brackets, of the several density estimators across 250 trials: ordinary smoothing spline (ORD), corrected smoothing spline (COR) and kernel density estimator (KDE). The length of the sampling interval is $\tau$ and $n$ is the sample size.}
		\label{tabla_mediana}
	\end{table}
	
	
	\section{Real data illustrations}
	\label{sec:realdata}
	
	Two datasets are now introduced to illustrate the practical performance of the smoothing spline density estimator compared to the kernel density estimator. The first dataset contains quasar luminosities, as considered in the seminal paper \cite{quasars}. The second dataset comprises ages at diagnosis of Parkinson's disease for the late-onset group considered in \cite{parkinson}. These two datasets, which suffer from double truncation, are available within the \texttt{DTDA} R package \cite{dtda}. The density estimates are computed as indicated in Section \ref{sec:sims}. Pointwise confidence limits for the density at 95\% level are obtained from the simple bootstrap that resamples with replacement from the original dataset $\{(U_i,V_i,X_i)\}_{i=1}^n$; for this, 250 bootstrap replicates are used.
	
	\subsection{Quasars data}
	
	This example, introduced by \cite{quasars} and included in the \texttt{DTDA} package, \texttt{Quasars} object, records the luminosity of observed quasars. The original dataset comprises quadruplets $(Z_i, m_i, U_i, V_i)$, where $Z_i$ is the redshift and $m_i$ is the apparent magnitude of the $i$th quasar. These two measures are transformed according to the cosmological model assumed (see \cite{quasars} for details), obtaining an adjusted luminosity measure $X_i=t(Z_i,m_i)$ in the log-scale. Due to experimental constraints, the distribution of each adjusted log-luminosity $X_i$ is truncated to a known interval $[U_i,V_i]$. Specifically, quasars with apparent magnitude above a certain threshold were too dim to yield dependent redshifts and right-truncation limit $V_i$ was obtained as a transformation of said threshold; the left-truncation limit $U_i$ was enforced in order to avoid confusion with non-quasar astronomical objects. Thus, only those observations with $X_i\in[U_i,V_i]$ are recorded, totalling $n=210$ triplets $(U_i,V_i,X_i)$. For this dataset the observation interval length $\tau_i=V_i-U_i$ is not constant.
	
\begin{figure}[H]
		\centering
		\includegraphics[width=\textwidth]{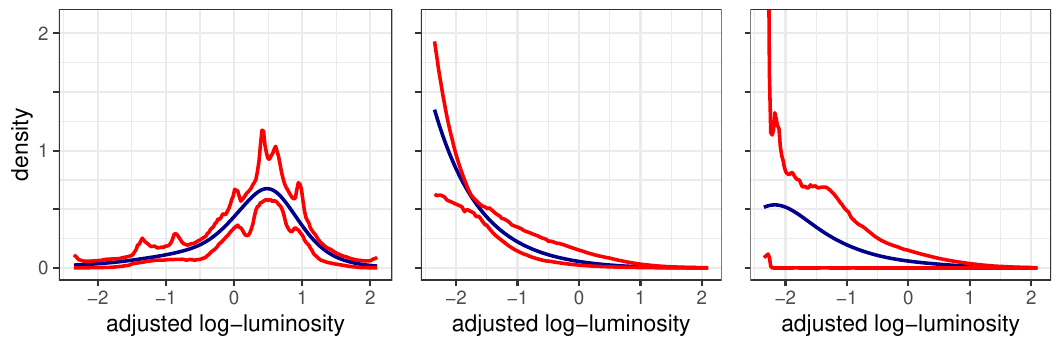}
		\caption{Density estimation results for the \texttt{Quasars} data obtained by three methods: ordinary smoothing spline (left), corrected smoothing spline (middle) and kernel density estimation (right). Represented in blue is the density estimate of the real data; in red, the 2.5\% top and bottom percentiles from 250 bootstrap estimates.}
		\label{panel_quasars}
	\end{figure}

	The results from the three different density estimation methods considered in this paper are displayed in Figure~\ref{panel_quasars}. It can be seen that the bias introduced by the truncation setting is substantial, causing the corrected smoothing spline to yield very different results from the ordinary spline. Although very few quasars at the lower end of the luminosity scale were recorded in the dataset, this is because of the empirical difficulty in observing them; after correcting for bias, most of the probability mass is concentrated at low-luminosity quasars. The kernel density estimates also detect and correct this bias, although the confidence limits are wider than those of the smoothing spline estimate. In addition, the kernel density estimator suffers from boundary effects at the left tail. All of these observations are in agreement with the results of the simulation study in Section \ref{sec:sims}.
	

	\subsection{Parkinson's disease data}
	
	\cite{parkinson} used these data to investigate the association between two candidate single nucleotide polymorphisms (SNPs) and age of onset of Parkinson's disease (PD). To avoid survival-related bias, the study was restricted to patients whose DNA samples were collected within 8 years of disease onset. As a result, variable $X$, years of age at onset of PD, is doubly truncated by the age at DNA sampling $V$, and $U=V-8$. This leads to an interval sampling setting with a subject-specific sampling interval. The case studies were divided into two groups: an early-onset group, comprising $n=99$ patients aged 35 to 55 at onset, and a late-onset group, consisting of $n=100$ patients aged 63 to 87. Here we consider the late-onset group, which is available from \texttt{DTDA} package, \texttt{PDlate} object.
	
\begin{figure}[h]
		\centering
		\includegraphics[width=\textwidth]{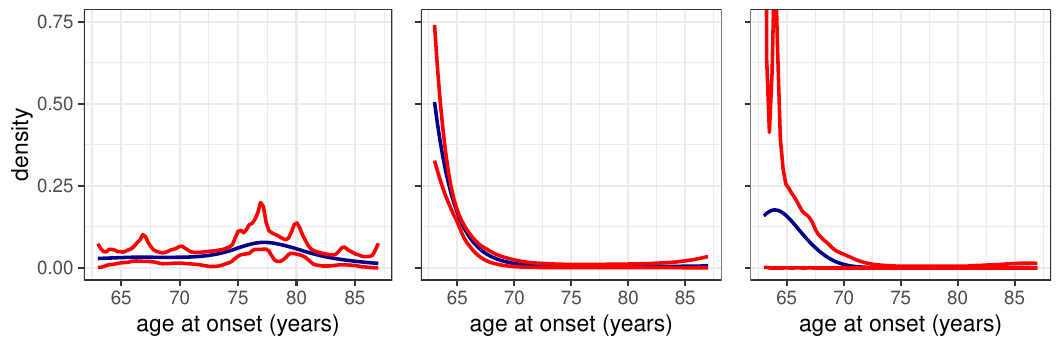}
		\caption{Density estimation results of the \texttt{PDlate} data obtained by the three methods: ordinary smoothing spline (left), corrected smoothing spline (middle) and kernel density estimation (right). Represented in blue is the density estimate of the real data; in red, the 2.5\% top and bottom percentiles from 250 bootstrap estimates.}
		\label{panel_PDlate}
	\end{figure}

	Figure~\ref{panel_PDlate} displays the density estimates obtained from the several methods. Compared with the ordinary smoothing spline, the bias correction performed by the modified smoothing spline estimator shifts the probability mass to the left. This is because the sampling probability is extremely low for ages below 70 (see \cite[Example 2.1.8]{dt.libro}). Interestingly, the corrected smoothing spline estimate exhibits small variance across the bootstrap replicates, in sharp contrast to the wide confidence limits of the KDE for ages below 70. As in Figure~\ref{panel_quasars}, some boundary effects for the kernel density estimate are also observed at the left tail.

	
\section{Main conclusions and discussion}
\label{sec:discussion}

In this paper smoothing spline density estimation for doubly truncated data has been investigated. From a theoretical viewpoint, validity of the proposed estimator can be guaranteed from the general theoretical framework corresponding to smoothing splines with biased data (see \cite{gu}). Practical implementation of the method has been performed through a proper modification of the \texttt{ssden} function within the package \texttt{gss}. Such modification was needed since \texttt{ssden} does not allow for three arguments in the bias function, which is the situation appearing under double truncation. This modified code is available from the authors upon request.

Simulation results have shown that the proposed method performs consistently, approaching the target as the sample size increases. Generally, although not always, the correction for double truncation results in a larger estimation variance; however, it is critical to remove the systematic bias of the ordinary smoothing spline estimator, which may arise from the biased sampling. Compared to other nonparametric estimation strategies such as kernel smoothing, the smoothing spline method is superior in our simulated settings. Two factors are important for this. First, kernel smoothing may suffer from a large variability due to the noise of the NPMLE $F_n$, which is convolved with the kernel function; moreover, such NPMLE may degenerate due to non-uniqueness or non-existence issues, leading to inadmissible density estimates. This problem does not affect the smoothing spline estimator, which does not use the NPMLE. Second, boundary effects, which are often present in kernel density estimation but naturally avoided by smoothing splines, can lead to a poorer performance of the former approach. These factors have been illustrated and discussed in Sections \ref{sec:sims} and \ref{sec:realdata}.

One drawback of smoothing spline density estimation, compared to kernel smoothing, is that it involves a larger computational time. For illustration, for the dataset on quasar luminosities ($n=210$) the execution time of the adapted \texttt{ssden} function was $0.8$ seconds, compared with $0.16$ seconds for the \texttt{densityDT} function. This additional computational effort becomes particularly important when performing bootstrapping for a large number of Monte Carlo trials.

\section*{Acknowledgements}

Second author was supported by the Grant PID2023-148811NB-I00 from Agencia Estatal de Investigación
(Ministerio de Ciencia, Innovación y Universidades, Spanish Government). This paper has been presented as invited oral talk at the 26th International Conference on Computational Statistics (COMPSTAT2024) held in Giessen, Germany, August 27-30 2024, and at the 4th IMS International Conference on Statistics and Data Science (ICSDS2025) held in Seville, Spain, December 15-18 2025; authors acknowledge feedback from the attendees.

\end{document}